\documentclass[a4paper,11pt,oneside]{article}
\usepackage[T1]{fontenc}
\usepackage{graphicx}
\usepackage{amsmath}
\usepackage{psfrag}
\usepackage{color}

\newcommand{\ds}[1]{\displaystyle{#1}}

\newcommand{\bq}{\begin{quote}}
\newcommand{\eq}{\end{quote}}

\newcommand{\cl}{\centerline}

\newcommand{\vs}{\vskip0.3cm\noindent}

\newcommand{\dequ}{\begin{equation} \vspace{3mm}}
\newcommand{\fequ}{\vspace{3mm} \end{equation}}

\newcommand{\ba}{\begin{array}}
\newcommand{\ea}{\end{array}}
\newcommand{\bc}{\begin{center}}
\newcommand{\ec}{\end{center}}
\newcommand{\bi}{\begin{itemize}}
\newcommand{\ei}{\end{itemize}}
\newcommand{\ben}{\begin{enumerate}}
\newcommand{\een}{\end{enumerate}}

\newcommand{\bfi}{\begin{figure}[hbtp]}
\newcommand{\efi}{\end{figure}}

\newcommand{\beq}{\begin{equation}}
\newcommand{\eeq}{\end{equation}}
\newcommand{\beqar}{\begin{eqnarray}}
\newcommand{\eeqar}{\end{eqnarray}}

\renewcommand{\bar}{\overline}
\newcommand{\bit}{\begin{itemize}}
\newcommand{\eit}{\end{itemize}}

\newcommand{\w}{\omega}

\newcommand{\eps}{\varepsilon}

\renewcommand{\vs}{\vskip0.1cm\noindent}

\renewcommand{\bq}{\begin{quote}}
\renewcommand{\eq}{\end{quote}}

\newcommand{\tensna}{\bar{\bar{\nabla}}}

\renewcommand{\ds}{\displaystyle}

\renewcommand{\vec }{\overrightarrow}

\begin{document}

\cl{\bf Chaotic particle sedimentation in a rotating flow }
\cl{\bf with time-periodic strength.}
\vskip.3cm
\cl{ J. R. Angilella}

\vskip.3cm

\cl
{Nancy-Universit\'e, LAEGO, Ecole Nationale Supérieure de Géologie,}
 \cl
{rue du Doyen Roubault, 54501 Vand\oe uvre-les-Nancy, France}
\cl{Jean-Regis.Angilella@ensem.inpl-nancy.fr}


\vskip.5cm
\cl{\bf Abstract}
{ 
Particle sedimentation in the vicinity of a  fixed horizontal vortex with time-dependent
intensity  can be chaotic, provided gravity is sufficient to displace the 
{particle cloud} whilst the vortex is off or weak. This "stretch, sediment \& fold" mechanism is close to
the so-called blinking vortex effect, which is  responsible for
 chaotic transport of perfect tracers, except that in the present case the vortex motion
is replaced by gravitational settling. 
In the present work this phenomenon is analyzed for heavy Stokes particles  moving under the sole effect of gravity
and of a linear  drag. The vortex is taken to be  a fixed isolated point vortex the intensity of which varies under the effect of either boundary conditions or volume force.
When the unsteadiness of the vortex is weak and the free-fall velocity is of the order of the fluid velocity, 
and the particle response time is small,
the particle motion equation can be written asymptotically 
as a perturbed hamiltonian system the phase portrait of which displays
a homoclinic trajectory. A homoclinic bifurcation is therefore likely to occur,
and the contribution of particle inertia to the occurrence of
  this bifurcation is analyzed asymptotically by using  Melnikov's method.  

}

 
\vskip.5cm


\section{Introduction}

In contrast with the chaotic advection of perfect tracers, which has been a topic of great interest in the last decades (Arnold \cite{Arnold1965}, H\'enon \cite{Henon1966},
Aref \cite{Aref2002}, Ottino \cite{Ottino1989}),  little is known about the chaotic motion of inertial particles in laminar flows.
Chaotic advection is entirely contained within
an elementary equation of kinematics, that is :
\beq
\frac{d\vec X}{dt} = \vec V_f\Big(\vec X(t),t \Big),
\label{puretracers}
\eeq
where $\vec X(t)$ is the position of a perfect tracer advected by the velocity field
$\vec V_f$. This velocity field is not required to be  turbulent
for Lagrangian chaos to take place :   it can be laminar and entirely specified.
In particular, chaotic advection provides efficient mixing properties, and this is   one of the reasons why
this effect attracted the attention of the physics 
community.
Mixing is indeed
a key phenomenon in nature, and a challenge for engineers as soon as one has to mix 
large amounts of very viscous fluids, or small quantities of fluid in tiny domains.
Nevertheless, the problem ``starts rather than ends with the specification of
the velocity field'' (Ottino \cite{Ottino1990}), so that many analyses are still being performed to quantify the Lagrangian properties of various flows.
Among the mechanisms leading to chaotic advection, the {\sl blinking vortex} is
probably the  easiest to perform (Aref \cite{Aref1984}). Two key ingredients are required 
here : the differential rotation and the displacement of the vortex. The former ingredient
is responsible for stretching and the latter is required for folding to take place. Many
chaotic flows are based on this property and are described in several reviews 
\cite{Aref2002}$^,$ \cite{Ottino1989}.

\vskip.3cm

In the case where  the particle is not a pure tracer, surprising chaotic trajectories can also occur, even when the inclusion obeys a linear drag law.
For example, aerosols can have complex trajectories in elementary cellular or ABC
flows (Maxey \& Corrsin
\cite{Maxey1986} ; Wang , Maxey, Burton \& Stock \cite{Wang1992} ; Mac Laughlin 
\cite{McLaughlin1988} ;   Fung \cite{Fung1997} ; Tsega, Michaelides \& Eschenazi  \cite{Tsega01} ; Rubin, Jones \& Maxey \cite{Rubin1995}). 
 The complex trajectories reported in these works
are   due to the  non-uniformity of
the flow, to the finite response time of the inclusion, and to gravity if any.

The goal of the present work is to show that 
heavy particles with small response time but 
non-negligible terminal velocities can have  chaotic trajectories under the combined effect
of a time-periodic differential rotation (which creates stretching of a particle cloud) and
of gravity (which prepares the particle cloud to undergo folding). In particular, we will consider flows
where unsteadiness does not suffice to induce Lagrangian chaos, like two-dimensional flows of the form :
\beq
\vec V_f(\vec X,t) = \vec V_f^0(\vec X)\, (1 + \eps \sin \w t)
\label{flowperturb}
\eeq
with  $\vec V_f^0(\vec X)$ corresponding to a horizontal vortex. Indeed, one can check that
the dynamical system (\ref{puretracers}) is not chaotic in this case, and that fluid points
go to and fro along the streamlines of the vortex. Clearly, unsteadiness does not produce
chaos because the vortex does not move : not only should we switch the vortex off sometimes,
but also should we light it up somewhere else to produce chaos according to the
 blinking vortex mechanism. However, if gravity is sufficient to displace the 
{particle cloud}
whilst the vortex is off (or weak), one could expect chaos according to some kind 
of ``gravity-induced blinking vortex''. Note, however, that the key 
role of gravity in this scenario could also be played
 by electrostatic forces, or  swimming (if particles are bacteria or plankton), to name
but a few examples. 
\vs
In the following we show that heavy Stokes particles can undergo such a mechanism.
The dynamics of inertial particles is much more complicated than the one
of tracers,
since one has to solve for both the equations of the flow induced by the inclusion, and the
motion equations of the inclusion. Nevertheless, significant simplifications
arise when the flow induced by the inclusion is a quasi-steady creeping flow.
In the case of tiny heavy particles (e.g. aerosols) carried by a fluid with
infinite extent one often writes
\beq
{\vec{\ddot X}_p} = \frac{\vec V_T}{\tau_p} + \frac{1}{\tau_p} \Big( \vec V_f(\vec X_p,t) -  \vec{\dot X}_p  \Big),
\label{eqmvt}
\eeq
where $\vec X_p(t)$ is the particle position, $\tau_p$ denotes
 its response time, 
and $\vec V_T$ is the terminal velocity 
of the inclusion in the very same  fluid at rest. 
This is the simplest equation for non-ideal tracers,
which requires the particle Stokes and Reynolds numbers to be much smaller than unity,
and brownian diffusion to be negligible.
In the present paper we will assume that (\ref{eqmvt}) is
non-dimensionalized by the typical length scale and velocity of $\vec V_f$ respectively,
and that :
$$
\tau_p  {\ll 1} \quad \mbox{and} \quad V_T = |\vec V_T| = O(1).
$$
The former condition can be thought of as a consequence of the fact that the
viscous time scale over the particle radius is much smaller than convective flow time scales. The
latter manifests non-negligible sedimentation effects. 
If in addition we assume $\eps \ll 1$, the particle motion equation contains two independent 
small parameters,
namely $\eps$ and $\tau_p$. 
Classical asymptotic expansions \cite{Maxey1987} of the form  
$
\vec{\dot X}_p = \vec V_T + \vec V_f^0 
 + \eps \vec V^{1} +   O(\eps^2)
$
enable one to write the particle motion equation as a three-dimensional non-autonomous
dynamical system :
\beq
\vec{\dot X}_p = \vec V_T + \vec V_f^0(\vec X_p) 
+ {\eps
\Big( \vec V_f^1(\vec X_p,t) -  k\,\tensna \vec V_f^0 . (\vec V_f^0 +\vec V_T)
\Big)}
+ O(\eps^2),
\label{eqmvt2ddl}
\eeq
where we have set
  $\tau_p = k\,\eps$, with $k$ held fixed as $\eps \to 0$. 
Because the flow is 2D and $div\,(\vec V_T + \vec V_f^0) = 0$, Eq. (\ref{eqmvt2ddl}) is a perturbed 
hamiltonian system with one and a half degree-of-freedom. The phase portrait of the
unperturbed system ($\eps = 0$) can display homoclinic or heteroclinic trajectories,
which are key-ingredients of chaos for such system. 
Figure \ref{SketchTraj}(b)
 shows 
such trajectories for particles moving in the vicinity of a point vortex the streamlines of
which are sketched in Fig. \ref{SketchTraj} (a).
The
 homoclinic  trajectory (dashed line)
 links the saddle point where the fluid velocity $\vec V_f^0(\vec X_p)$
balances the terminal velocity $\vec V_T$.
(Throughout this paper we have chosen  $\vec V_T = -V_T\vec e_y$, $V_T > 0$).
 Such particle trajectories have 
often been observed in particle-laden flows. For example, trajectories like Fig. 
\ref{SketchTraj}(b) have been investigated by Davila \& Hunt \cite{Davila2001}.
\vs
The perturbations contained in the $O(\eps)$ terms of Eq. (\ref{eqmvt2ddl}) can have 
several tremendous effects on the particle dynamics. In particular, 
a homoclinic bifurcation can occur, leading to chaotic particle settling
or trapping. The purpose of the present paper is to investigate under 
which conditions such a bifurcation could occur.

\section{Asymptotic analysis in the vicinity of the separatrix}

  The basic vortical flow investigated in this paper is the point vortex   : 
\beq
\vec V_f^0(\vec x) = r \Omega(r) \vec e_\theta, \quad \mbox{with} \quad \Omega(r) = \Omega_0(r/R_0)^{-2}
\label{v0}
\eeq
This velocity field is set non-dimensional in the following by using $R_0$ as length units and $1/\Omega_0$ as time units.  The streamfunction $\psi^0$ of this non-dimensional flow therefore reads :
$\ds{
\psi^0(\vec X_p(t)) = -\frac{1}{2} Log |\vec X_p |^2 .
}$
To leading order $O(\eps^0)$ the   particle dynamics reads :
\beq
\vec{\dot X}_p = \vec V_T + \vec V_f^0(\vec X_p) 
\label{eps0}
\eeq
and the corresponding trajectories therefore correspond to the iso-values of the hamiltonian :
$$
H(x,y) = \psi^0(x,y) + x V_T 
$$
and are sketched in figure  \ref{SketchTraj} (b). In order to investigate separatrix splitting under the effect of the $O(\eps^1)$ terms it is necessary to solve analytically the leading-order motion (\ref{eps0})  
with $\vec X_p(\pm \infty) = A$ (saddle point). To our knowledge, even for the simple flow considered here, this cannot be done. To be precise one cannot obtain a simple solution which would make the Melnikov integral easy to calculate analytically.
However, one can obtain semi-analytical results by rescaling the variables. Indeed, we set :
$$
\vec X_p(t) = \frac{1}{V_T} \vec Y(\tau) \quad \mbox{with} \quad \tau = t V_T^2
$$
and notice that the velocity field investigated here satisfies :
$$
\vec V_f^0(\vec X_p) = V_T \vec V_f^0(\vec Y).
$$
 The particle dynamics therefore reads  : 
$$
\frac{d \vec Y}{d\tau} = \vec V_f^0(\vec Y) -  \vec e_y 
$$
\beq
+\eps \left[ \frac{1}{V_T} \vec V_f^1\left( \frac{1}{V_T} \vec Y(\tau),\frac{\tau}{V_T^2}\right)
-k V_T^2 \,\tensna_Y \vec V_f^0(\vec Y) . (\vec V_f^0(\vec Y) -  \vec e_y )
   \right]
\label{dynrenorm}
\eeq
The leading order dynamical equation is now independent of $V_T$, and the corresponding phase portrait in the $\vec Y$ plane is similar to the one of figure  \ref{SketchTraj} (b), with the saddle point located at $(1,0)$  (figure \ref{PoincY}(a)). Let $\vec Y_0(\tau)$ be a solution of the leading-order dynamics :
\beq
  \dot{\vec Y}_0   = \vec V_f^0(\vec Y_0) -  \vec e_y , \quad \vec Y_0(\pm \infty) = (1,0)
\label{y0}
\eeq
(the dot upon a $\vec Y$ indicating a derivation with respect to $\tau$) which satisfies, accordingly :
$$
\ddot{\vec Y}_0 =\tensna_Y \vec V_f^0  . ( \vec V_f^0(\vec Y_0) -  \vec e_y ).
$$
When the $O(\eps)$ terms are taken into account, and because the perturbation of the rescaled system is time-periodic (with period $T_Y = T V_T^2$), one usually considers the Poincar\'e section (or stroboscopic map) $\vec Y(\tau_0+n T_Y)$, with $n=1,2,...$ Because the Poincar\'e section of the unperturbed system has a hyperbolic point at $(1,0)$, the Poincar\'e section of the perturbed system will have a hyperbolic point of the same kind (saddle) in the vicinity of $(1,0)$, provided $\eps$ is small enough. An unstable (resp. stable) invariant manifold $W^u$ (resp.$W^s$ ) will therefore exist in the vicinity of the hyperbolic point. If these two manifolds intersect transversally, they will have an infinity of such intersection points.
The non-dissipative (area preserving) character of the $O(\eps^0)$ system will induce huge stretching, and folding will inevitably follow.
The phase portrait in the vicinity of the vortex will
take the form sketched in figure \ref{PoincY}(b) which drastically differs from the non-chaotic case.  Particle sedimentation will then be chaotic. Such intersection points can be detected by making use of the classical Melnikov method (see for
example Guckenheimer \& Holmes \cite{GH01}) which consists in calculating the dot product $d(\tau_0) = \vec{EF}.\vec N$, where $\vec N$ is the normal to the homoclinic trajectory at $\vec Y_0(0)$ with $(\dot{\vec Y}_0,\vec N,\vec e_z)$ right-handed, $F$ (resp. E) is the intersection between this normal and  $W^u$ (resp. $W^s$ ). If the two manifolds intersect transversally then $d(\tau_0)$ will have simple zeros as $\tau_0$ varies. If  $d(\tau_0)$ remains strictly negative, then the relative position of the two manifolds will look like the one sketched in figure  \ref{PoincY}(c) : $W^u$ lying outside and $W^s$ inside. In this case the particles located outside the cell will fall regularly without being catched into the cell, whereas the particles initially located inside the cell will spiral out and exit the cell.
To order $O(\eps)$, the Melnikov "distance"  $d(\tau_0)$  is proportional to the Melnikov function\cite{GH01} : 
$$
M_{\vec Y}(\tau_0) = \int_{-\infty}^\infty \dot{\vec Y}_0(\tau) \wedge \vec V_f^0(\vec Y_0(\tau))  \sin \left( \frac{\w}{V_T^2}(\tau+\tau_0) \right)\, d\tau
-k V_T^2 \int_{-\infty}^\infty \dot{\vec Y}_0(\tau) \wedge \ddot{\vec Y}_0(\tau)\, d\tau
$$
where $\tau_0$ is the starting time of the Poincar\'e section of the dynamical system.
By writing   $\vec Y_0 = (\xi(\tau),\eta(\tau))$ and expanding the sine function, and noticing that $\dot\xi$  is an odd function, we are led to :
$$
M_{\vec Y}(\tau_0) = \cos \frac{\w \tau_0}{V_T^2} \int_{-\infty}^\infty  \dot\xi(\tau) \sin  \frac{\w \tau}{V_T^2} \, d\tau - k V_T^2 \gamma
$$
where 
$$
\gamma = \int_{-\infty}^\infty  (\dot\xi\, \ddot\eta - \dot\eta \, \ddot\xi)    \, d\tau
$$
is a purely numerical constant. As noticed above, both $\xi(\tau)$ and $ \eta(\tau)$ are unknown, but they are purely numerical functions which can be determined from a numerical solution of equation  (\ref{y0}). Finally, the Melnikov function of the homoclinic trajectory in the $\vec Y$ plane reads
\beq
M_{\vec Y}(\tau_0) = 
\cos \left(\frac{\w}{V_T^2}\tau_0\right)
 F\left(\frac{\w}{V_T^2}\right) - k V_T^2  \gamma
\label{resMelnik}
\eeq
 where $F(s)$ is the sine transform of $\dot\xi$ and is a purely numerical function depending only on the shape of the initial vortex. Note that since $ \dot\xi\, \ddot\eta - \dot\eta \, \ddot\xi$ is proportional to the curvature of the homoclinic trajectory, the constant part $- k V_T^2 \gamma $ clearly manifests the contribution of a centrifugal effect due to particle inertia, as already observed for the onset of chaos in Stommel cells\cite{Angilella2007}.

Because $F(0) = 0$, the steady case $\w=0$ is  
straightforward :
$
M_{\vec Y}(\tau_0) =  - k V_T^2  \gamma < 0.
$
This means that the manifold $W^u$ remains outside the cell, whereas $W^s$ remains in the inner side  (like in figure \ref{PoincY}(c)) : they will not intersect, and any particle  released outside the cell will go down without penetrating into the cell. 
Also, as already mentioned, particles released inside the cell will spiral out.
We recover the fact that permanent suspension does not exist for such inertial particles in our flow, as already noticed by Wang \& Maxey \cite{Maxey1986}, Rubin, Jones \& Maxey \cite{Rubin1995} for other flows.

In the unsteady case $\w > 0$, the Melnikov function  has simple zeros if the amplitude of the oscillating term is larger than the constant term.
The criterion for the appearance of  chaotic particle sedimentation is therefore :
\beq
F\left(\frac{\w}{V_T^2}\right) > k V_T^2  \gamma .
\label{critere}
\eeq
\vs
By making use of a numerical   algorithm to solve Eq. (\ref{y0}) we obtain $\gamma \approx 45.8$, together with $F(s)$ which is plotted on figure \ref{Fxfig}. This function has a peak value
$\max_s F(s) \approx 1.14$, so that no chaotic motion is expected to occur, under the present hypotheses, if (say)
$$
k \gamma V_T^2 \ge 1.15
$$
In this case the Melnikov function remains strictly negative whatever the frequency of the perturbation : the particle motion is  always regular.
For $k \gamma V_T^2 < 1.14$ a homoclinic bifurcation can occur, provided $\omega/V_T^2$ lies in an appropriate range ("chaotic window") like the one shown in figure \ref{Fxfig}.  The  particle motion is therefore highly affected by the  oscillations of the vortex.

\section{Comparison with  numerical solutions}

Figure \ref{NuagesXfig} shows the evolution of a particle cloud initially released inside the cell
obtained by solving  numerically  the   particle motion equation  (\ref{eqmvt}) with $\eps=0.1$. The terminal velocity of the particles is $V_T= 0.42$ and the frequency of the perturbation is $\w=4 V_T^2\simeq 0.7$. In addition $k=0.1$, so that
$k \gamma V_T^2 \simeq 0.8$ : according to figure  \ref{Fxfig}    chaos is likely to occur in the vicinity of the separatrix.
 We indeed observe that 
the particle cloud is folded and stretched. Figure \ref{DeuxTrajs} shows two typical  particle trajectories, initially released outside the cell for $k \gamma V_T^2 \simeq 0.8$. In the case $\w = 5 V_T^2$ the motion is chaotic and the spatial length of the trajectory is larger than in the non-chaotic case  ($\w = 11 V_T^2$) because the particle is captured into the cell and spins there for a while. This property does not imply that the system is chaotic, but it is used in the following to detect the occurrence of chaos in a more systematic way, as proposed by Ziemniak \& Jung \cite{Ziemniak1995}, and as done in a previous paper \cite{Angilella2007}.

Indeed, to check the predictive power of formula (\ref{critere}), we have run a set of computations where 1000 particles are released slightly above the cell, with $\eps=0.2$ and $\tau=0.02$. The trajectory of each particle $p$ is then calculated by introducing a random phase shift $a_p$ in the flow perturbation (simply replace $\w t$ by $\w t +a_p$ in equation (\ref{flowperturb})).
The calculation is stopped when the particle reaches a fixed bottom,  below the cell. We then calculate the centred averaged path length :
$$
\theta(\w) = \frac{L(\w) - L(\infty)}{L(\infty)}
$$
where $L(\w)$ is the average particle path. The quantity $\theta(\w)$ is plotted in figure \ref{PathLength} for four values of $V_T$.
In addition, we have plotted the amplitude of the Melnikov function minus its constant part, that is $F(\w/V_T^2) - k \gamma V_T^2$, the positiveness of which implies chaotic particle motion.
It appears that when $F(\w/V_T^2) - k \gamma V_T^2 < 0$ the average particle path length is closed to zero (i.e. all the particle paths have roughly the same length), as expected if the two manifolds do not intersect (like in figure \ref{PoincY}(c)). In contrast,
a soon as $F(\w/V_T^2) - k \gamma V_T^2 > 0$ the average particle path length increases, and this manifests the fact that some particles  have penetrated into the cell, as a consequence of the homoclinic bifurcation.

\section{Discussion}

 The calculations presented in this note show that a fixed vortex is sufficient to induce chaotic particle motion, under the sole effect of gravity and of the unsteadiness of the vortex. We have chosen to consider a fixed vortex with time-dependent intensity. One could argue that this choice is not realistic, since the intensity (circulation) of vortices is known to remain constant unless viscosity affects it (Kelvin's theorem). This is why we assumed that some appropriate boundary conditions, or an appropriate volumic force, was present to sustain the whole picture. Our goal being to show that gravity and {\sl unsteady differential rotation } are sufficient ingredients to induce chaotic particle settling. Note also that the detailed shape of the vortex  (here a rotation rate decaying like $1/r^2$, where $r$ is the distance to the vortex centre) might not be of major importance, and that other decaying rotation rates  could also lead to chaotic particle motion.

In the present analysis, gravity plays a significant role since it is responsible  for the appearance of a homoclinic trajectory in the leading-order dynamics.
Like for particle settling in the vicinity of upward streamlines\cite{Angilella2007}, particle inertia is opposed to the appearance of chaos, because of centrifugal effects, and tends to maintain the two invariant manifolds $W^s$ and $W^u$ away from each other.
The flow unsteadiness, in contrast, tends to make these  manifolds intersect.

To check the predictive power of the Melnikov analysis we have computed the average length of particle paths, as proposed by
Ziemniak \& Jung \cite{Ziemniak1995} in the framework of fluid points trajectories in the wake of a cylindrical obstacle. These authors observed that the 
probability to find a trajectory whose path length increase is larger than some value $s$,  
is an exponential function of $s$. It would be therefore of interest to check whether such exponential distributions are visible also in the present case. Clearly,
these statistics are linked to the area of the lobes (i.e.   subsets of the phase space located between $W^u$ and $W^s$, as shown in grey on figure \ref{PoincY}(b)), which can be calculated from the Melnikov function. In addition, lobe dynamics can be used to calculate other global quantities like particle flux accross the separatrices (see for example Rom-Kedar, Leonard \& Wiggins \cite{RomKedar1990},
Balasuriya \cite{Balasuriya2005}). A detailed analysis of such integral quantities would therefore be of interest to characterize more precisely inertial particle transport in the chaotic regime. Further studies on  this topic should be of interest. 

In the absence of particle inertia ($\tau_p=0$) the particle dynamics is always chaotic (just set $k=0$ in the Melnikov function (\ref{resMelnik})) : this is a pure hamiltonian chaos, in the sense that   the complete   system is hamiltonian here, and very close to chaotic advection of perfect tracers. The Poincar\'e sections in this case are very classical, and are shown in figure \ref{SecPoinc} in the case $V_T=1$. As expected, KAM curves are visible outside the stochastic layer close to the homoclinic trajectory, indicating that some particles could be trapped for a while, and these curves are destroyed as the unsteadiness of the flow increases.

The present mechanism could be applied to mixing devices in chemical engineering processes, where one could leave sedimentation act between every two stiring periods. To our knowledge, detailed mathematical analyses devoted to such devices have not been published so far. In another context, 
the stretch, sediment and fold mechanism could play a non-negligible role in the mixing of plankton (or the mixing of any other "particle") in the upper ocean.
Indeed,
under the combined effect of settling and of unsteady wind-induced (or temperature gradient induced) rotating flows, patches of sedimenting particles could perhaps be mixed efficiently. This point needs further discussions.

The main conclusion of the present work is that an unsteady differential rotation is sufficient to induce chaotic heavy particle settling  provided the still-fluid terminal velocity of the inclusion is close enough to the flow velocity, and particle inertia is small enough. (In contrast, fluid point trajectories are very regular here, as fluid points go to and fro along portions of circle.)
The mechanism, which can be called "stretch, sediment \& fold", is sketched in figure \ref{SketchBlink} (for the sake of clarity,  large amplitude oscillations are assumed there) : a particle cloud is  stretched by a vortex, then the vortex weakens and the cloud sediments, then the vortex starts again and the cloud is folded, and so on. This elementary mechanism  is different from the one investigated by  Vilela \& Motter \cite{Vilela2007} with two blinking vortex-sources.
Indeed, in the system investigated by these  authors
 gravity is not a key ingredient for the appearance of chaos.
 The particle spirals out around the vortex-sources, until it reaches a limit cycle and remains suspended permanently. On this attractor the spiraling time of the inclusion is of the order of the period of the blinking. In our single-vortex case, it is  the sedimentation time scale which has to match the blinking period. Moreover, these authors show that a cascade of period doublings occurs as the particle inertia decreases, leading to a strange attractor. 
 In the present paper no attractor can be observed, since our system is reduced to a perturbed hamiltonian system. It could   therefore be of interest to check whether, once   this simplification is removed, a permanent suspension could appear in the vicinity of a single fixed singularity  with time-periodic strength, and gravity. Further analyses should clarify this point.



\vskip1cm
\vs
{\Large\bf Acknowledgement}
\vs
The author wishes to thank R.D. Vilela and A.E. Motter for fruitful discussions at the M. Planck Institute for the Physics of Complex Systems in Dresden.
Support from the network ``Syst\`emes Dynamiques Chaotiques''
 of INPL (Nancy, France) is gratefully acknowledged.  

 


\newpage
\bfi
\cl{\includegraphics[height=7cm]{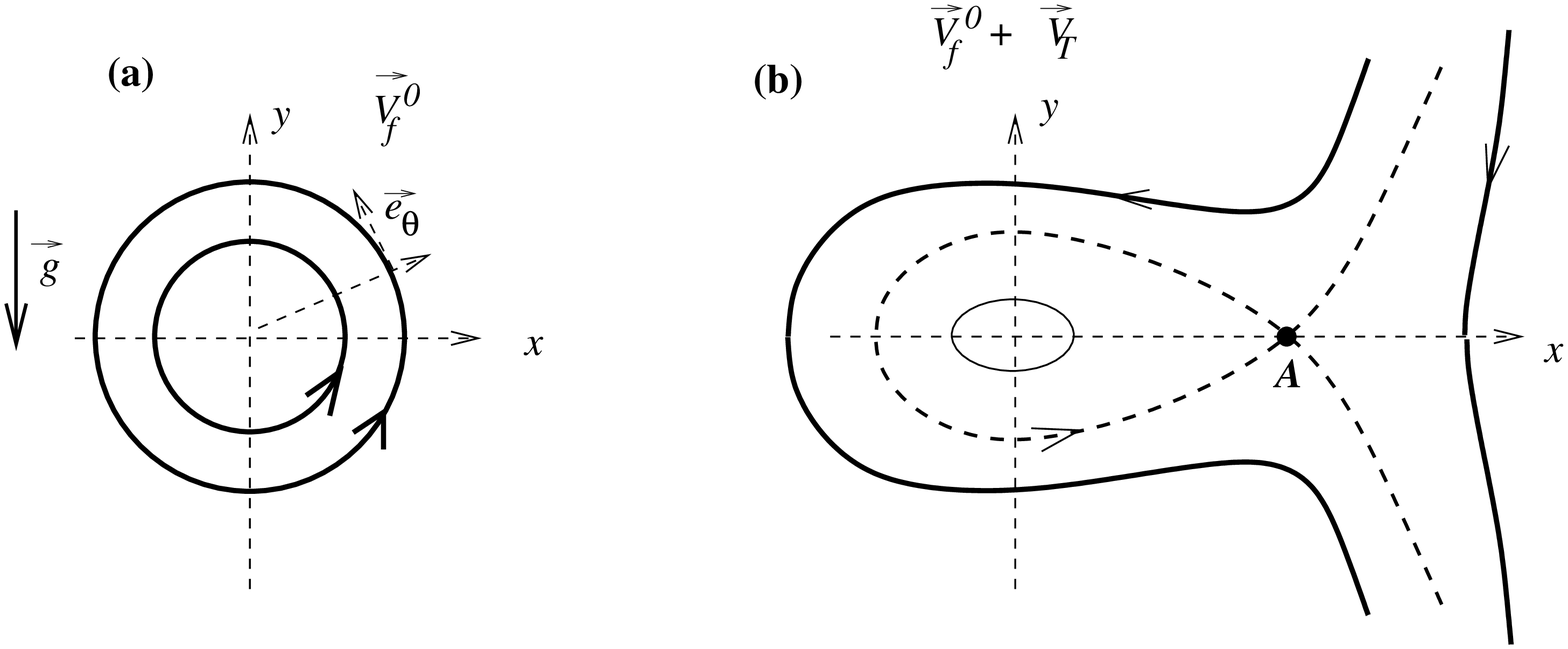}}
\caption{Sketch of the streamlines of a horizontal vortex  (a) and of  particle trajectories (to leading order $\eps^0$) in the vicinity of this vortex (b). The
 homoclinic  trajectory (dashed line)  is attached to the saddle point $A$ where the fluid velocity balances the terminal velocity.} 
\label{SketchTraj}
\efi

\newpage

\bfi
\cl{\includegraphics[height=10.cm]{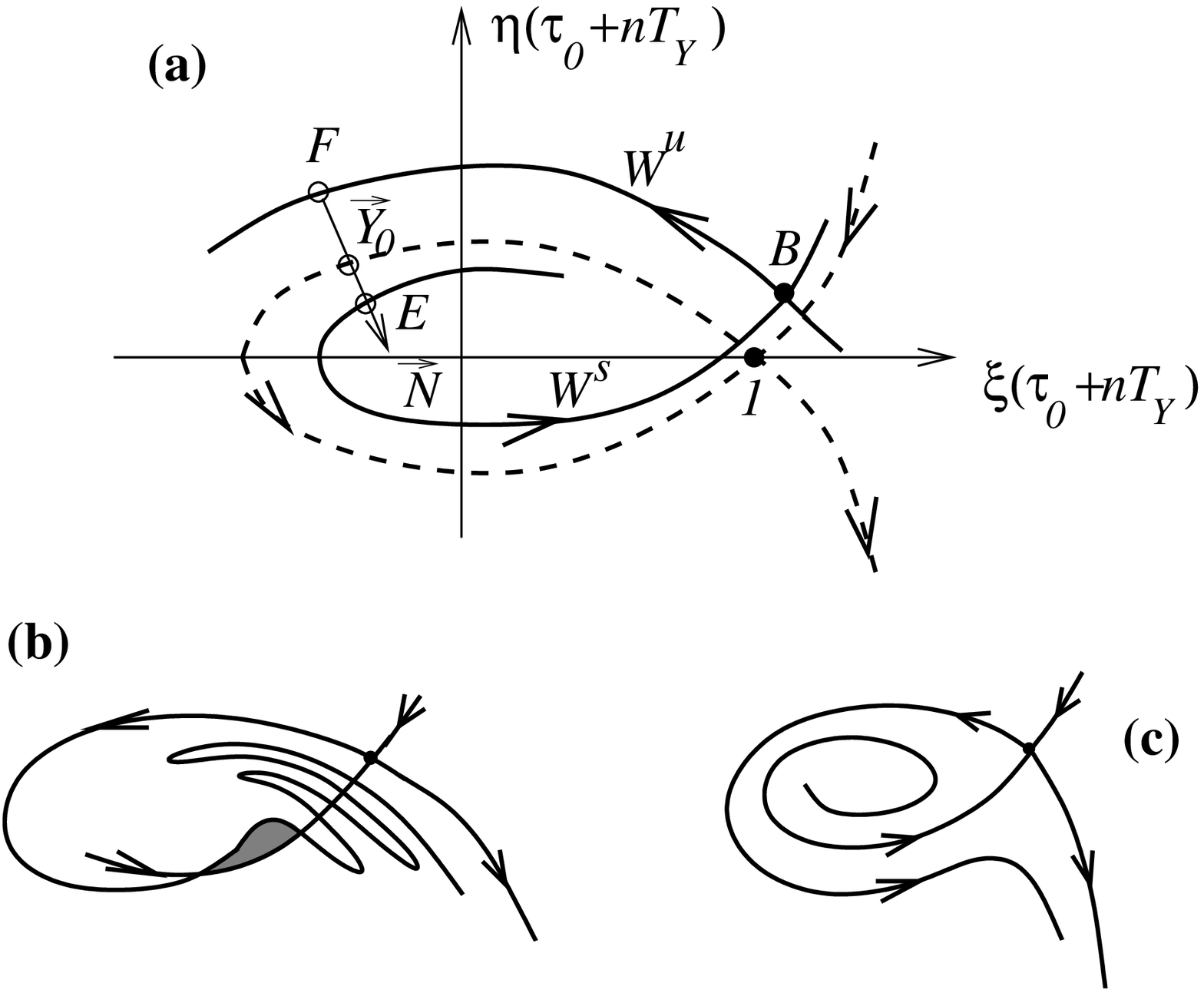}}
 
\caption{Sketch of the Poincar\'e section of the perturbed renormalized system. In case (b) the manifolds $W^s$ and $W^u$ intersect, leading to chaotic particle trajectories. In case (c)  the two manifolds do not intersect : particles released inside the cell will spiral out, and those released outside will go round the cell.  }
\label{PoincY}
\efi

\newpage

\bfi
\cl{\includegraphics[height=9.cm]{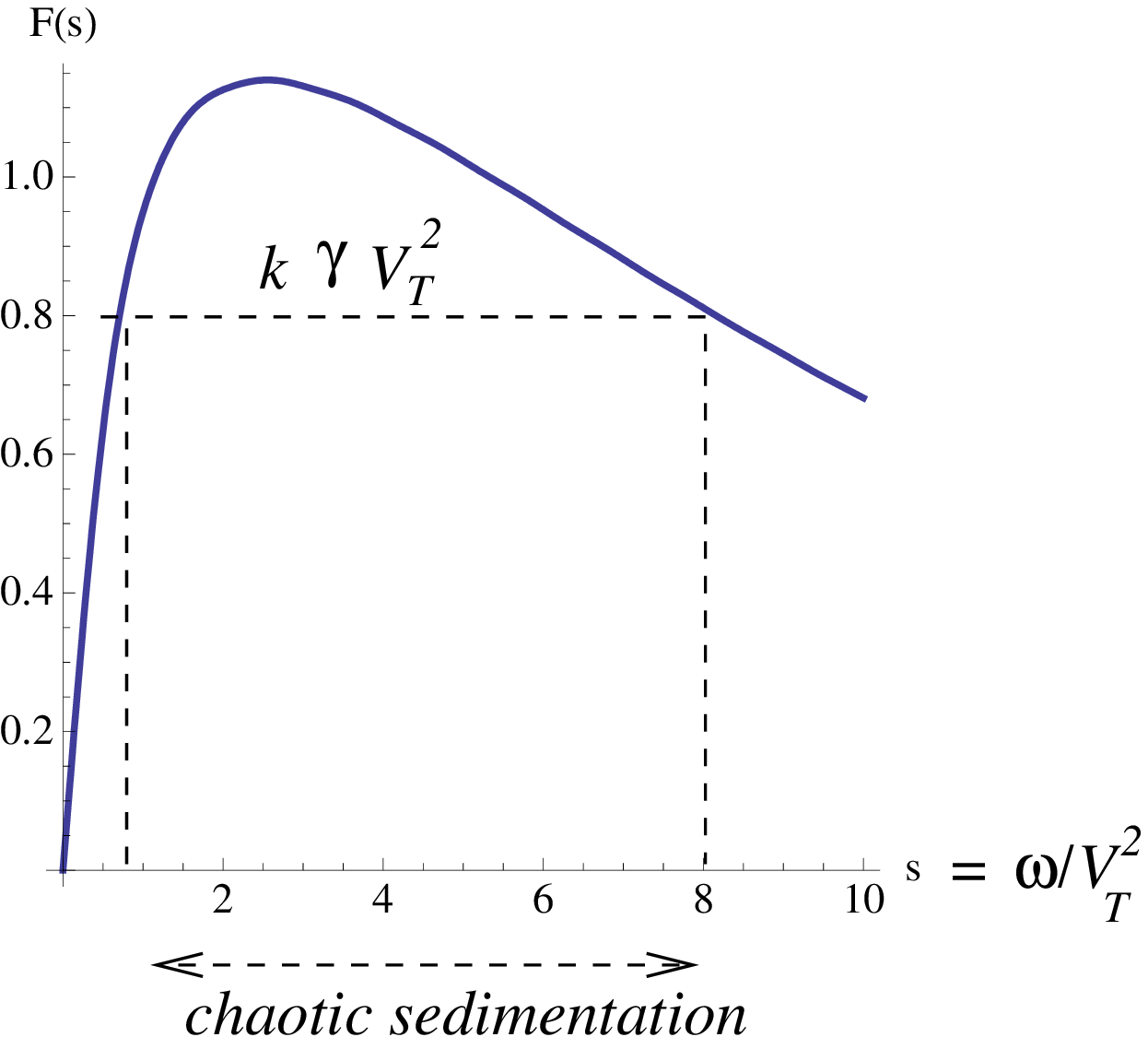}}
\caption{Plot of the amplitude of the Melnikov function, obtained by solving numerically the particle path over the homoclinic trajectory of the renormalized system.}
\label{Fxfig}
\efi

\newpage

\bfi
\cl{\includegraphics[height=10cm]{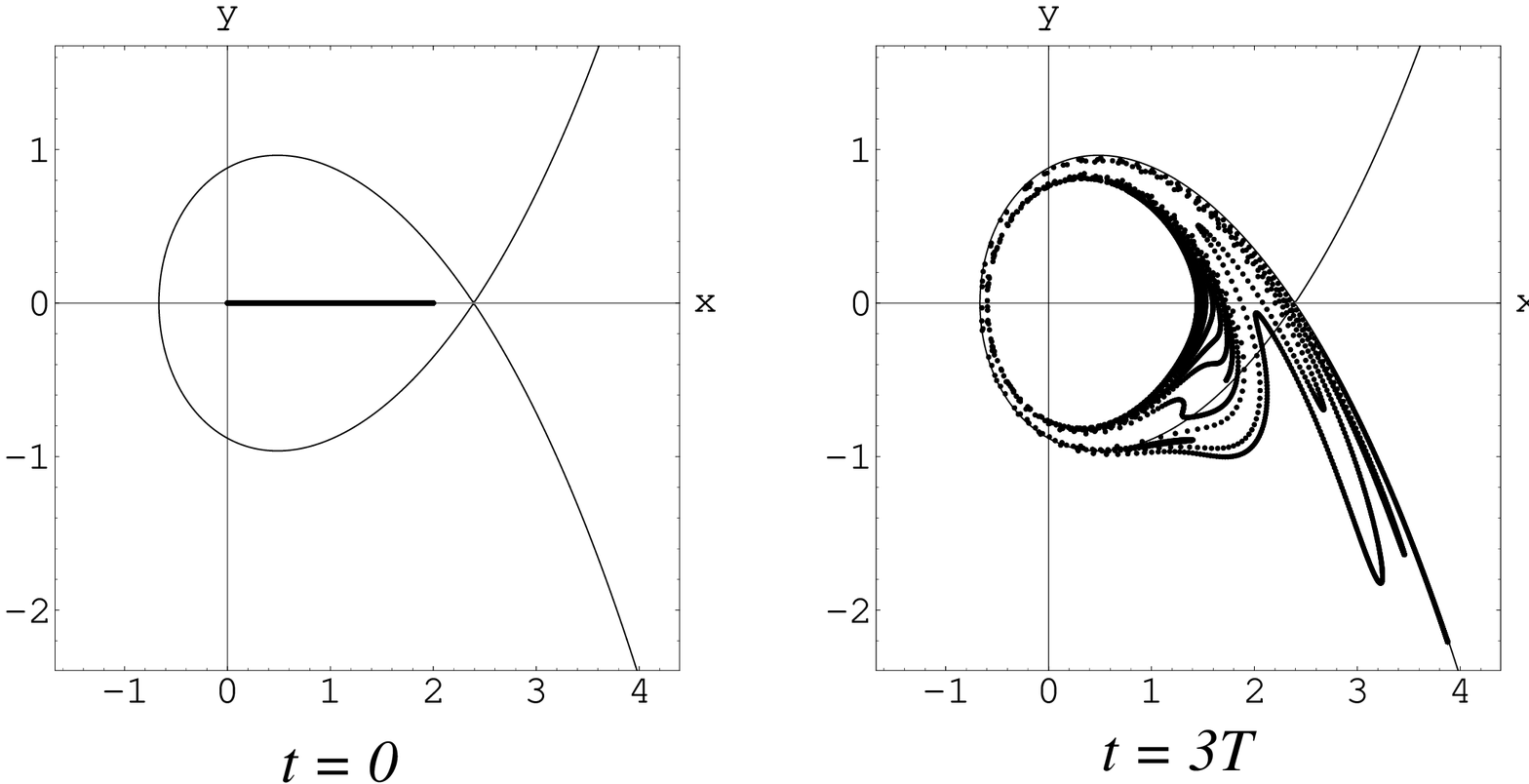}}
\caption{Evolution of a particle cloud initially released inside the cell, when $\w=4 V_T^2$ and
$k \gamma V_T^2 \simeq 0.8$   (chaotic case).}
\label{NuagesXfig}
\efi

%
%

\bfi
\cl{\includegraphics[height=6.5cm]{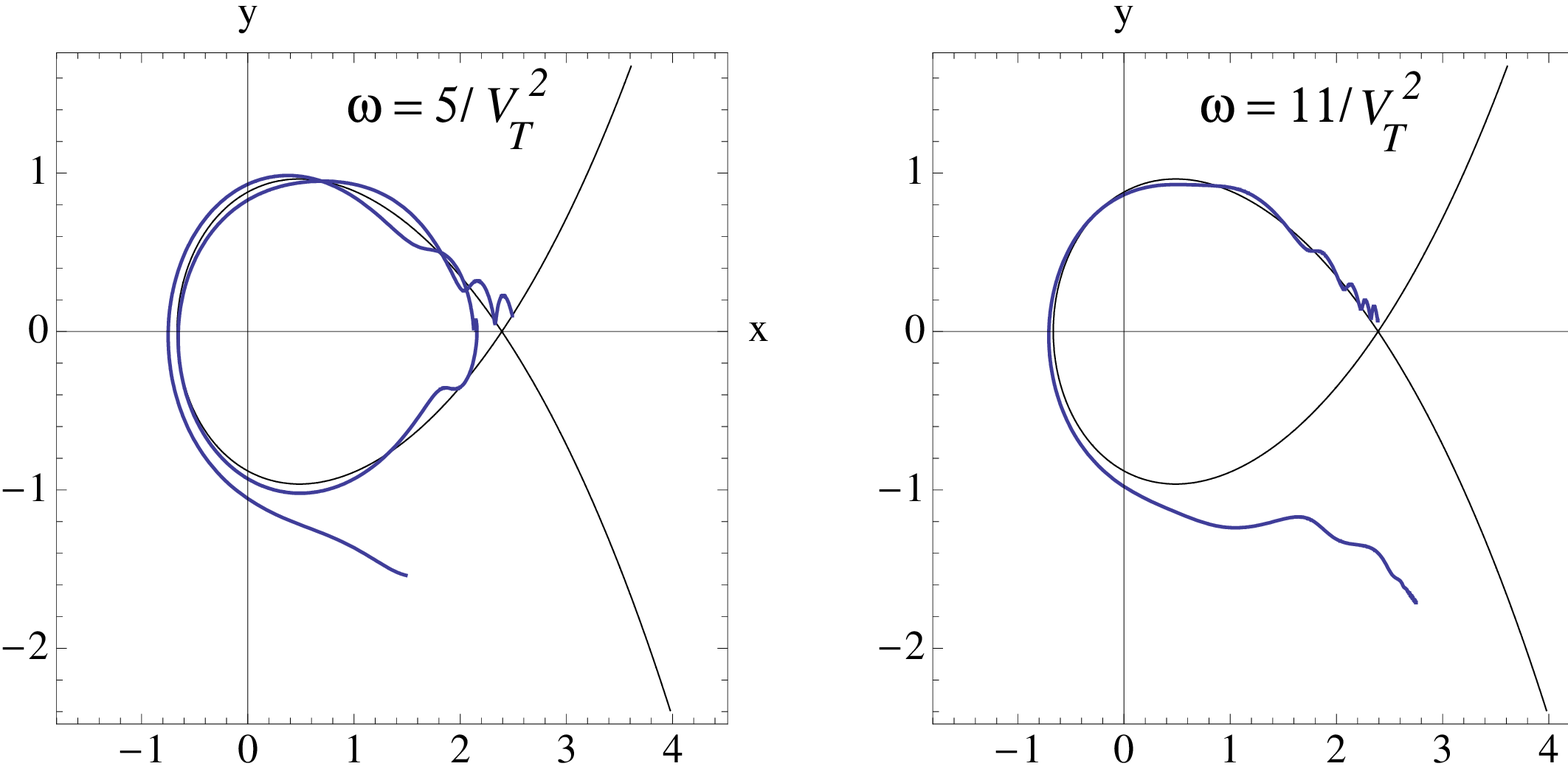}}
\caption{Typical particle paths in the chaotic (left) and non-chaotic (right) cases, with $k \gamma V_T^2 = 0.8$.}
\label{DeuxTrajs}
\efi

\newpage 

\bfi
\cl{\includegraphics[height=12cm]{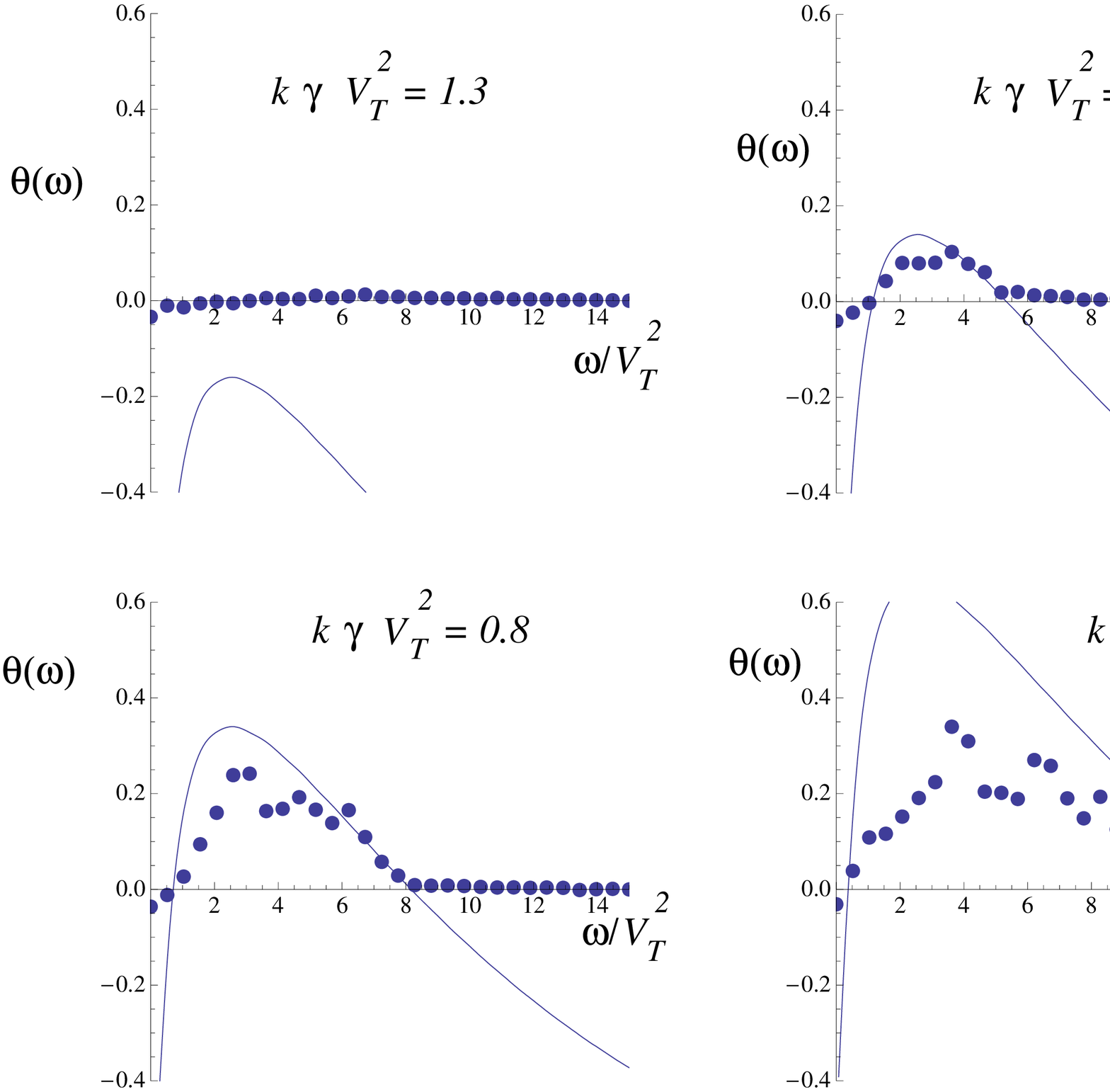}}
\caption{Plot of the centred average path length ($\theta(\w)$, black dots), together with the amplitude of the Melnikov function minus its constant part  (solid line,  formula (\ref{critere})). As soon as this line is above zero, the separatrix gets broken, some particles therefore penetrate into the cell and $\theta(\w)$ increases.  }
\label{PathLength}
\efi

\bfi
\cl{\includegraphics[height=15cm]{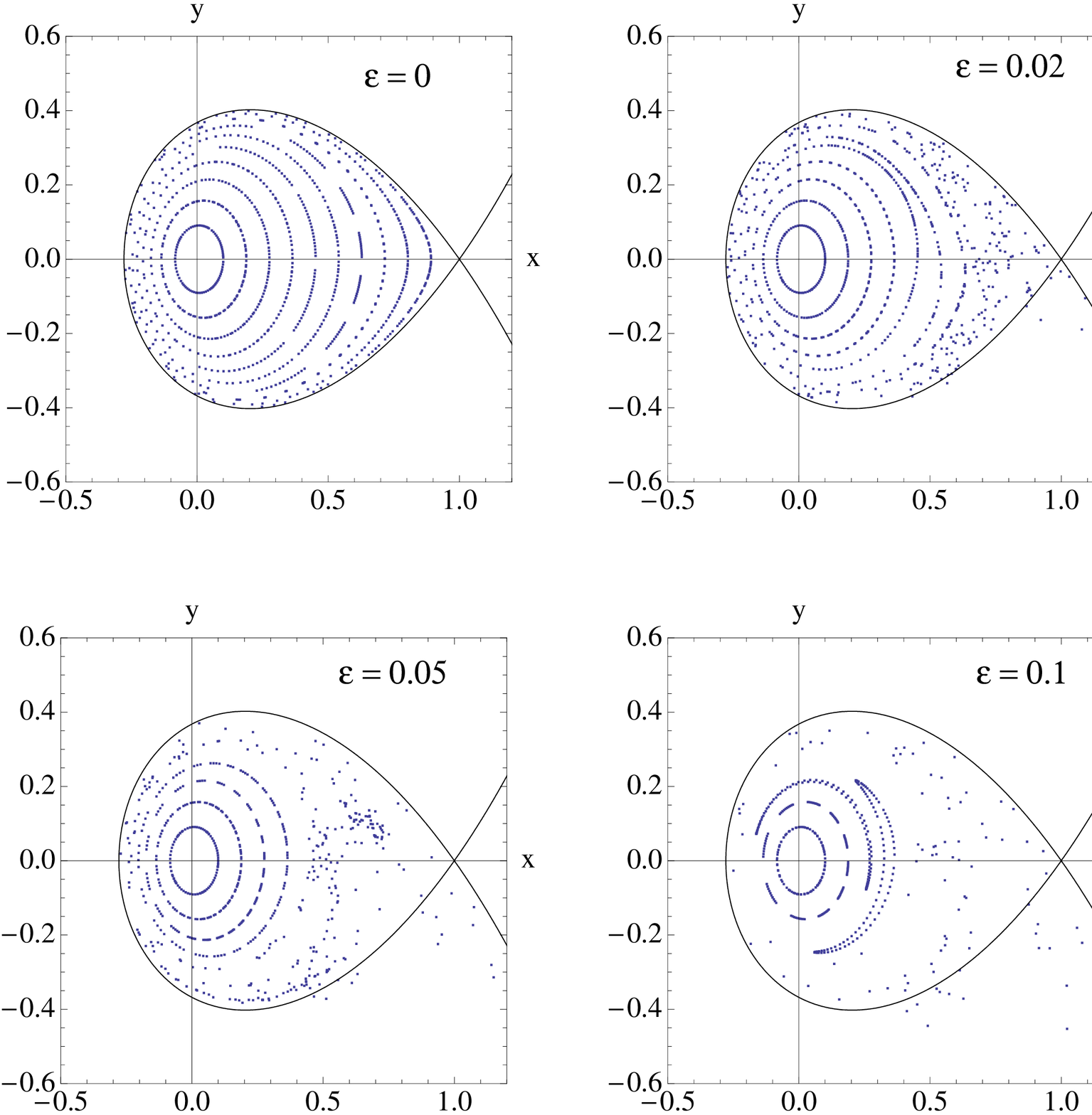}}
\caption{Plot of 10 Poincar\'e sections for $\tau_p=0$ and $V_T=1$. Because  $k=0$ the   criterion (\ref{critere}) obtained from Melnikov's analysis is always fulfilled, so that the stochastic layer in the vicinity of the separatrix always exists (except, of course, in the steady case $\eps=0$). The plots show the destruction of KAM tori as the amplitude of the perturbation increases.}
\label{SecPoinc}
\efi

\bfi
\cl{\includegraphics[height=5cm]{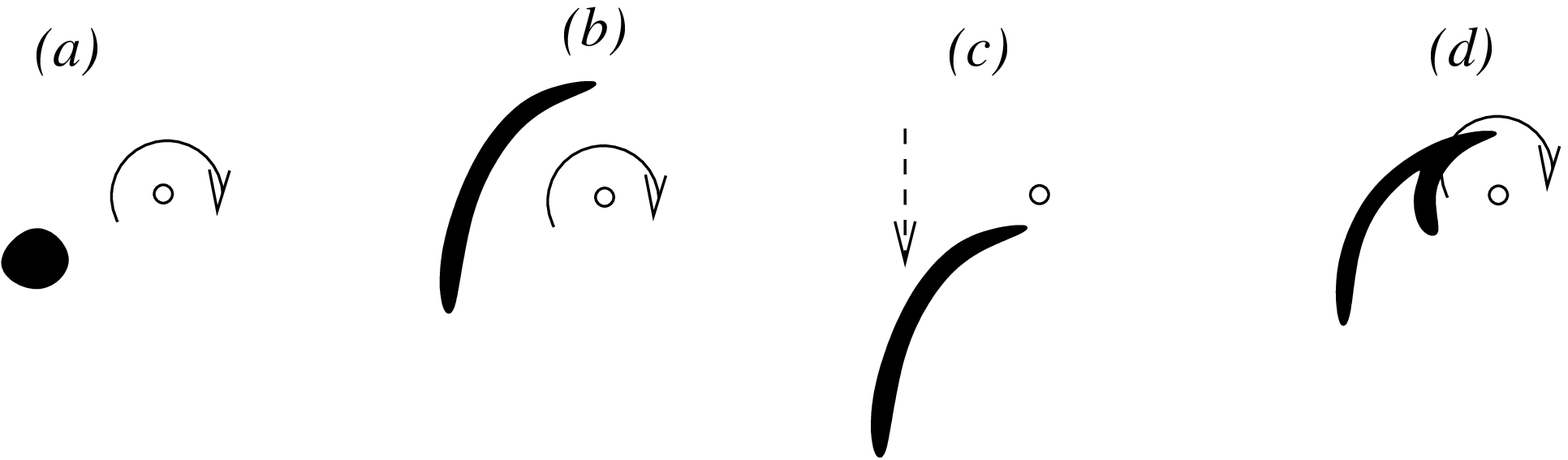}}
\caption{Sketch of the gravity-induced blinking vortex effect, for large amplitude oscillations. 
Sketch (a) shows the initial particle cloud (black) and the vortex (white circle).
In step (b) the particle cloud is stretched due to differential rotation. Then the vortex weakens and the cloud sediments (c). When the differential rotation restarts the cloud is folded.}
\label{SketchBlink}
\efi

\end{document}